\documentstyle[epsf,twocolumn,prd,aps,floats]{revtex}
\begin{document}

\draft
%
%
\newcommand{\nc}{\newcommand}
\nc{\bea}{\begin{eqnarray}}
\nc{\eea}{\end{eqnarray}}
\nc{\beq}{\begin{equation}}
\nc{\eeq}{\end{equation}}
\nc{\bi}{\begin{itemize}}
\nc{\ei}{\end{itemize}}
\nc{\la}[1]{\label{#1}}
\nc{\half}{\frac{1}{2}}

\nc{\IH}{{}^1{\rm H}}
\nc{\D}{{\rm D}}
\nc{\EH}{{}^3{\rm H}}
\nc{\EHe}{{}^3{\rm He}}
\nc{\UHe}{{}^4{\rm He}}
\nc{\GLi}{{}^6{\rm Li}}
\nc{\ZLi}{{}^7{\rm Li}}
\nc{\ZBe}{{}^7{\rm Be}}
\nc{\DH}{{\rm D}/{\rm H}}
\nc{\EHeH}{^3{\rm He}/{\rm H}}
\nc{\ZLiH}{{}^7{\rm Li}/{\rm H}}
\nc{\lgZLiH}{\log_{10}({}^7{\rm Li}/{\rm H})}
\nc{\et}{\eta_{10}}
\nc{\etl}{\eta_{\rm low}}
\nc{\eth}{\eta_{\rm high}}
\nc{\ropt}{r_{\rm opt}}

\nc{\etal}{{\it et al.}}
\nc{\x}[1]{}
%
%
\nc{\AJ}[3]{{Astron.~J.\ }{{\bf #1}{, #2}{ (#3)}}}
\nc{\anap}[3]{{Astron.\ Astrophys.\ }{{\bf #1}{, #2}{ (#3)}}}
\nc{\ApJ}[3]{{Astrophys.~J.\ }{{\bf #1}{, #2}{ (#3)}}}
\nc{\apjl}[3]{{Astrophys.~J.\ Lett.\ }{{\bf #1}{, #2}{ (#3)}}}
\nc{\apjs}[3]{{Astrophys.~J.\ Supp.\ S.\ }{{\bf #1}{, #2}{ (#3)}}}
\nc{\app}[3]{{Astropart.\ Phys.\ }{{\bf #1}{, #2}{ (#3)}}}
\nc{\araa}[3]{{Ann.\ Rev.\ Astron.\ Astrophys.\ }{{\bf #1}{, #2}{ (#3)}}}
\nc{\arns}[3]{{Ann.\ Rev.\ Nucl.\ Sci.\ }{{\bf #1}{, #2}{ (#3)}}}
\nc{\arnps}[3]{{Ann.\ Rev.\ Nucl.\ and Part.\ Sci.\ }{{\bf #1}{, #2}{ (#3)}}}
\nc{\epj}[3]{{Eur.\ Phys.\ J.\ }{{\bf #1}{, #2}{ (#3)}}}
\nc{\MNRAS}[3]{{Mon.\ Not.\ R.\ Astron.\ Soc.\ }{{\bf #1}{, #2}{ (#3)}}}
\nc{\mpl}[3]{{Mod.\ Phys.\ Lett.\ }{{\bf #1}{, #2}{ (#3)}}}
\nc{\Nat}[3]{{Nature (London) }{{\bf #1}{, #2}{ (#3)}}}
\nc{\ncim}[3]{{Nuov.\ Cim.\ }{{\bf #1}{, #2}{ (#3)}}}
\nc{\nast}[3]{{New Astronomy }{{\bf #1}{, #2}{ (#3)}}}
\nc{\ndt}[3]{{Nuclear Data Tables\ }{{\bf #1}{, #2}{ (#3)}}}
\nc{\np}[3]{{Nucl.\ Phys.\ }{{\bf #1}{, #2}{ (#3)}}}
\nc{\pr}[3]{{Phys.\ Rev.\ }{{\bf #1}{, #2}{ (#3)}}}
\nc{\PRC}[3]{{Phys.\ Rev.\ C\ }{{\bf #1}{, #2}{ (#3)}}}
\nc{\PRD}[3]{{Phys.\ Rev.\ D\ }{{\bf #1}{, #2}{ (#3)}}}
\nc{\PRL}[3]{{Phys.\ Rev.\ Lett.\ }{{\bf #1}{, #2}{ (#3)}}}
\nc{\PL}[3]{{Phys.\ Lett.\ }{{\bf #1}{, #2}{ (#3)}}}
\nc{\prep}[3]{{Phys.\ Rep.\ }{{\bf #1}{, #2}{ (#3)}}}
\nc{\pla}[3]{{Plasma Phys.\ }{{\bf #1}{, #2}{ (#3)}}}
\nc{\RMP}[3]{{Rev.\ Mod.\ Phys.\ }{{\bf #1}{, #2}{ (#3)}}}
\nc{\rpp}[3]{{Rep.\ Prog.\ Phys.\ }{{\bf #1}{, #2}{ (#3)}}}
\nc{\zphysA}[3]{{Z.\ Phys.\ A }{{\bf #1}{, #2}{ (#3)}}}
\nc{\ibid}[3]{{\it ibid.\ }{{\bf #1}{, #2}{ (#3)}}}

\wideabs{
\title{Inhomogeneous Big Bang Nucleosynthesis and the High Baryon
Density Suggested by Boomerang and MAXIMA
}

\author{Hannu Kurki-Suonio\cite{mailh}}
\address{Helsinki Institute of Physics,
         P.O.Box 9, FIN-00014 University of Helsinki, Finland}

\author{Elina Sihvola\cite{maile}}
\address{Department of Physics, University of Helsinki,
         P.O.Box 9, FIN-00014 University of Helsinki, Finland}

\maketitle

\begin{abstract}
The recent Boomerang and MAXIMA data on the cosmic microwave
background suggest a large value for the baryonic matter density
of the universe, $\omega_b \sim 0.03$.  This density is larger
than allowed by standard big bang nucleosynthesis
theory and observations on the abundances of the light elements.
We explore here the possibility of accommodating this high density
in inhomogeneous big bang nucleosynthesis (IBBN).  It turns out that
in IBBN the observed $\D$ and $Y_p$ values are quite consistent
with this high density.  However, IBBN is not able to reduce the
$\ZLi$ yield by more than about a factor of two.
For IBBN to be the solution, one has
to accept that the $\ZLi$ plateau in population II halo stars
is depleted from the primordial abundance by at least a factor of two.

\end{abstract}

 }


%
%
\section{Introduction}

The most accurate way to estimate the average density of baryonic
matter in the universe, $\rho_b$, has for a long time been big
bang nucleosynthesis.
In standard big bang nucleosynthesis (SBBN)\cite{SBBN}
the calculated primordial yields of the light elements depend only
on the baryon-to-photon ratio $\eta \equiv n_b/n_\gamma$.
Future precision measurements of the fluctuations in the cosmic
microwave background (CMB) will provide another way of measuring
the baryon density.  The shape of the angular power spectrum of
these fluctuations will depend on a number of cosmological
parameters, among which is the baryon-to-photon ratio.  In this
context it is usually given as the baryonic contribution to the
critical density times the Hubble constant squared, $\omega_b
\equiv \Omega_bh^2$.

These two measurements rely on completely
different physics, and if they agree it is an important
confirmation that we have the right understanding of the early
universe.  Assuming that there was no significant entropy
production\cite{KT00} in the universe between BBN and recombination, the two
parameters $\omega_b$ and $\eta$ are related via the present
temperature of the CMB, $T = 2.725$ K, by
\beq
   \et = 274 \omega_b.
\eeq

Two recent balloon-borne experiments, Boomerang\cite{boom1} and
Maxima-1\cite{max1}, have now provided us with the first
measurements of the CMB angular power spectrum which are of sufficient
quality that an estimate of $\omega_b$ can be made from them.
The result\cite{boommax}, $\omega_b \sim 0.030$,
is significantly higher
than the SBBN result.  Indeed, SBBN clearly cannot accommodate
as high a baryon density as $\omega_b = 0.030$\cite{BoomSBBN}.

The SBBN yields for $\omega_b = 0.030$ are $Y_p = 0.251$,
$\DH = 1.7\times10^{-5}$, $\EHeH = 8.4\times10^{-6}$, and
$\ZLiH = 7.6\times10^{-10}$.  With the exception of $\ZLi$, the
uncertainty in these SBBN yields is much less than the uncertainty
in the primordial abundances derived from observations.

There is much debate about chemical evolution and systematic effects in
observations. For the primordial helium abundance there are two
competing estimates, the ``low $\UHe$''\cite{loHe}, $Y_p =
0.234\pm0.003$ and the ``high $\UHe$''\cite{hiHe}, $Y_p =
0.244\pm0.002$.  The difference is largely due to the
method of estimating the present abundances from the observed line intensities,
suggesting that the systematic uncertainty may be larger than
the $\pm0.005$ usually
assumed\cite{Olive00}.

Burles and Tytler \cite{Tytler} claim to have established the
primordial deuterium abundance  as $\DH =
(3.3\pm0.25)\times10^{-5}$ (``low D") based on Lyman-series absorption by high-redshift
clouds.  This is based on a detected low deuterium abundance in three such systems and
upper limits from others.  There remains one
such system, where a high deuterium abundance, $\DH \sim
2\times10^{-4}$ (``high D"), has been observed\cite{hiD}.  It may be that the
accuracy of such observations has been
overestimated\cite{mesoturb}.
We refer the reader to
recent reviews\cite{BBNreviews} for further discussion and adopt the observational
constraints $Y_p = 0.228$--$0.248$, $\DH =
2.9$--$4.0\times10^{-5}$\cite{Steigman}.  These lead to the SBBN
range $\omega_b = 0.004$--$0.021$ from $Y_p$ and $\omega_b = 0.018$--$0.022$ from $\DH$.
The ``low D'' of Burles and Tytler gives $\omega_b =
0.020\pm0.01$.

The ``Spite plateau''\cite{Spite} of $\ZLi$
abundance in population II halo stars provides us with
the best estimate of the primordial lithium abundance.  Bonifacio
and Molaro\cite{BoMo97} obtain a present lithium abundance
$\lgZLiH = -9.80\pm0.012\pm0.05$ for these stars, and argue against any
significant depletion from the primordial abundance,
based on the lack of dispersion in the data.
Pinsonneault \etal\cite{Pins} estimate a depletion factor of 0.2--0.4 dex
from rotationally induced mixing.
In a more recent study Ryan
\etal\cite{Ryan99} obtain a mean abundance $\lgZLiH = -9.88$
for the Spite plateau,
and argue that the narrow spread, less than 0.02 dex, limits depletion
by rotationally induced mixing to less than 0.1 dex.
Moreover, they observe a slight trend with
metallicity, suggesting a galactic contribution, so that the
primordial abundance could be lower than the observed abundance.
Based on this, Ryan \etal\cite{Ryan00} derive the range
$\lgZLiH = -10.04$\ldots$-9.72$ for the primordial abundance.
This very tight upper limit to primordial lithium is in conflict
with the ``low D'' estimate\cite{Tytler}, $\omega_b = 0.020$, which
corresponds to $\lgZLiH = -9.5$
in SBBN.

At present there seems to be a consensus\cite{Pins,VC,Deli},
that the primordial abundance cannot be
larger than the observed value by more than about 0.4 dex, so that
models producing a primordial lithium abundance above $\lgZLiH =
-9.4$, or $\ZLiH = 4\times10^{-10}$, would be ruled out.  This
gives an upper limit $\omega_b \leq 0.022$ in SBBN.

While the upper limit to $\omega_b$ from $Y_p$ is rather soft,
the deuterium and lithium abundances are in clear conflict with
$\omega_b = 0.03$.

This CMB measurement of $\omega_b$ is rather
preliminary and not very accurate, and thus it would be premature to discard
the SBBN estimate.  Since the CMB measurement is a simultaneous fit of many
cosmological parameters, the lack of prominence of the second peak in the power
spectrum could have another explanation.  Suggestions include a significant
tilt\cite{TZboomWSP00}, an isocurvature component\cite{EKV00},
or a feature\cite{Barriga} in the primordial
perturbation spectrum, and topological defects\cite{topo}.

Future satellite experiments, MAP\cite{MAP} and
Planck\cite{Planck}, will give us a better CMB determination
of $\omega_b$.  If the situation persists and the CMB value
of $\omega_b$ turns out to be $\omega_b \sim 0.03$ or higher,
one has to consider abandoning SBBN in favor of
non-standard BBN\cite{NSBBN}.
There are many modifications suggested to standard BBN, some of
which can accommodate a higher baryon density\cite{NSBBNhighden}.

We consider here one such possible modification, inhomogeneous big
bang nucleosynthesis (IBBN).  In IBBN one assumes that the baryon
density was inhomogeneous during BBN.  This inhomogeneity could be
an initial condition resulting from unknown physics in the early
universe, or it could be the result of some physical event, like a
phase transition, occurring before BBN.

The nature of IBBN depends on the distance scale of the
inhomogeneity.  Especially interesting is the case, where this
distance scale is of the order of the neutron diffusion length
during BBN\cite{earlyIBBN}.  Then neutron diffusion will lead to an inhomogeneity
in the neutron-to-proton ratio.  If the density contrast between
high- and low-density regions is large enough, this ratio can
exceed unity in the low-density region.  In this case IBBN leads
to a decrease in the $\UHe$ yield and an increase in the $\D$
yield, favoring larger $\omega_b$.  However, the
$\ZLi$ yield is more likely to increase than decrease from
the SBBN result for the
same average $\omega_b$.

In this paper we focus on the question, whether the high value
$\omega_b = 0.030$, corresponding to $\et = 8.22$, would be
acceptable in IBBN.  There have been many papers on IBBN\cite{earlyIBBN}.
However, because of the large number of parameters in the IBBN
model, the parameter space has not been covered thoroughly in
them.  Focusing on one particular value of $\et$ we reduce the
dimension of the parameter space, allowing us to search the space
of the remaining parameters more completely.  We can thus give a
more specific answer to the above question.

\section{Results}

We have updated our IBBN code\cite{KKS99} with the latest
compilation of reaction rates\cite{compilation}.
The new reaction rates have little impact on the isotope yields.
The largest
effect is on $\ZLi$ production, which is reduced by 5\%.
Vangioni-Flam \etal\cite{uncertainty} estimate the uncertainty in $\ZLi$ yield due to
reaction rates to be 30\%. The actual uncertainty may be even larger, since
their method of error estimation allows error compensation.

We assume a simple initial density profile, so that there is
a high density region, with one baryon density ($\eth$), and a low-density
region with another baryon density ($\etl$).  Diffusion will then smooth
out this initial density profile.  We assume a spherically
symmetric geometry, allowing us to study two different geometric
configurations:  1) centrally condensed (CC), where we have a
spherical high density region surrounded by a low density region,
and 2) spherical shell (SS), where we have a spherical shell of
high baryon density inside which we have low baryon density.
The first case approximates a situation where the high density regions
have a rather compact shape, with a low surface-to-volume ratio,
the second case a situation where
the high density regions have a more planar geometry, with a large
surface-to-volume ratio.

Once the average baryon density $\eta$ is fixed, we have three
remaining parameters, the density contrast $R \equiv \eth/\etl$,
the volume fraction $f_v$ of the high-density region, and the
distance scale $r$.  There is an ``optimum value'' $\ropt$ of the
distance scale, related to the neutron diffusion scale, where the
effect of the neutron excess is maximized. For very high density
contrasts, $f_vR \gtrsim 100$ the results become independent of
$R$.  For a more detailed discussion, see Ref.~\cite{KKS99}.

\begin{figure}[tbh]
\epsfysize=6.5cm
\epsffile{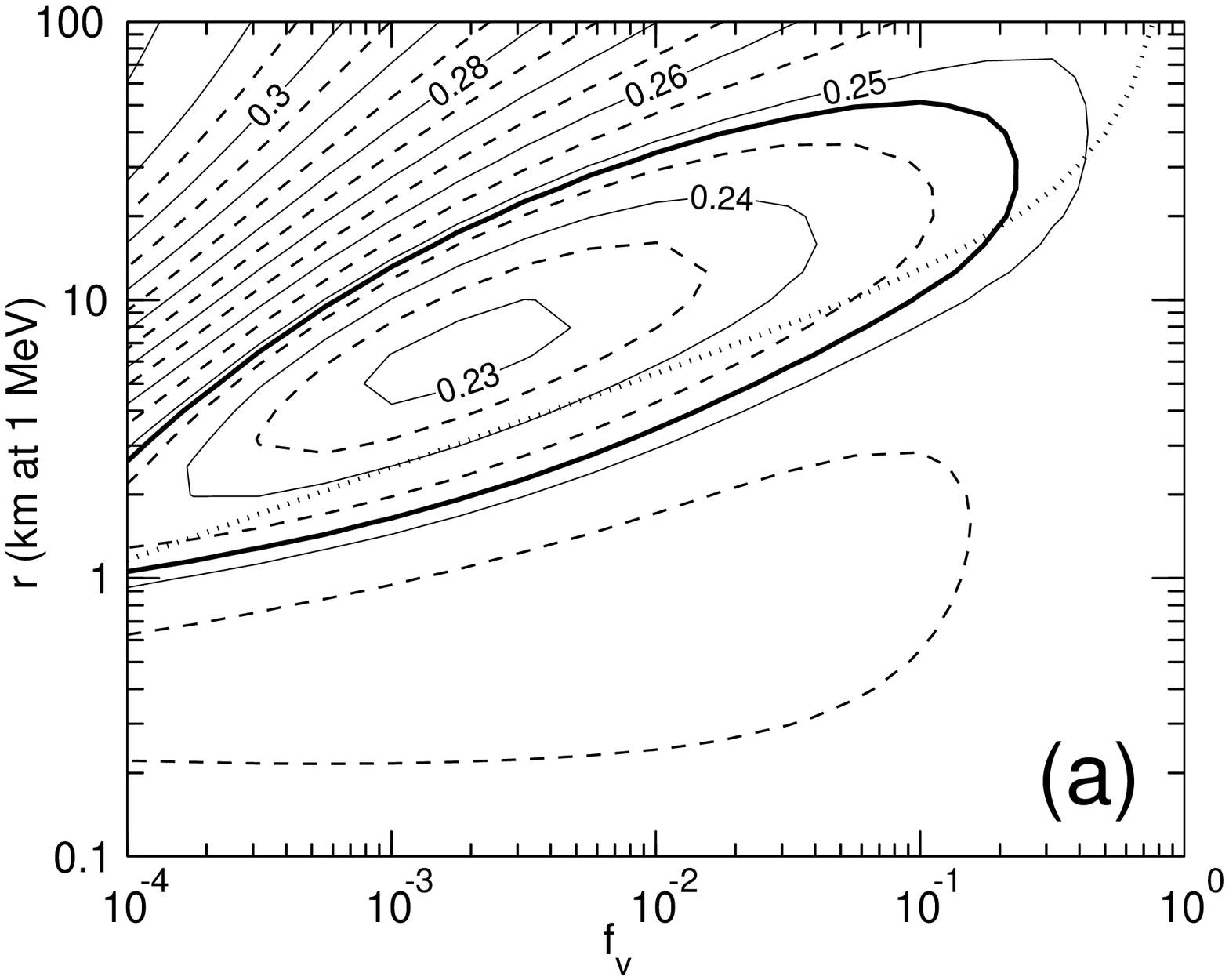}
\epsfysize=6.5cm
\epsffile{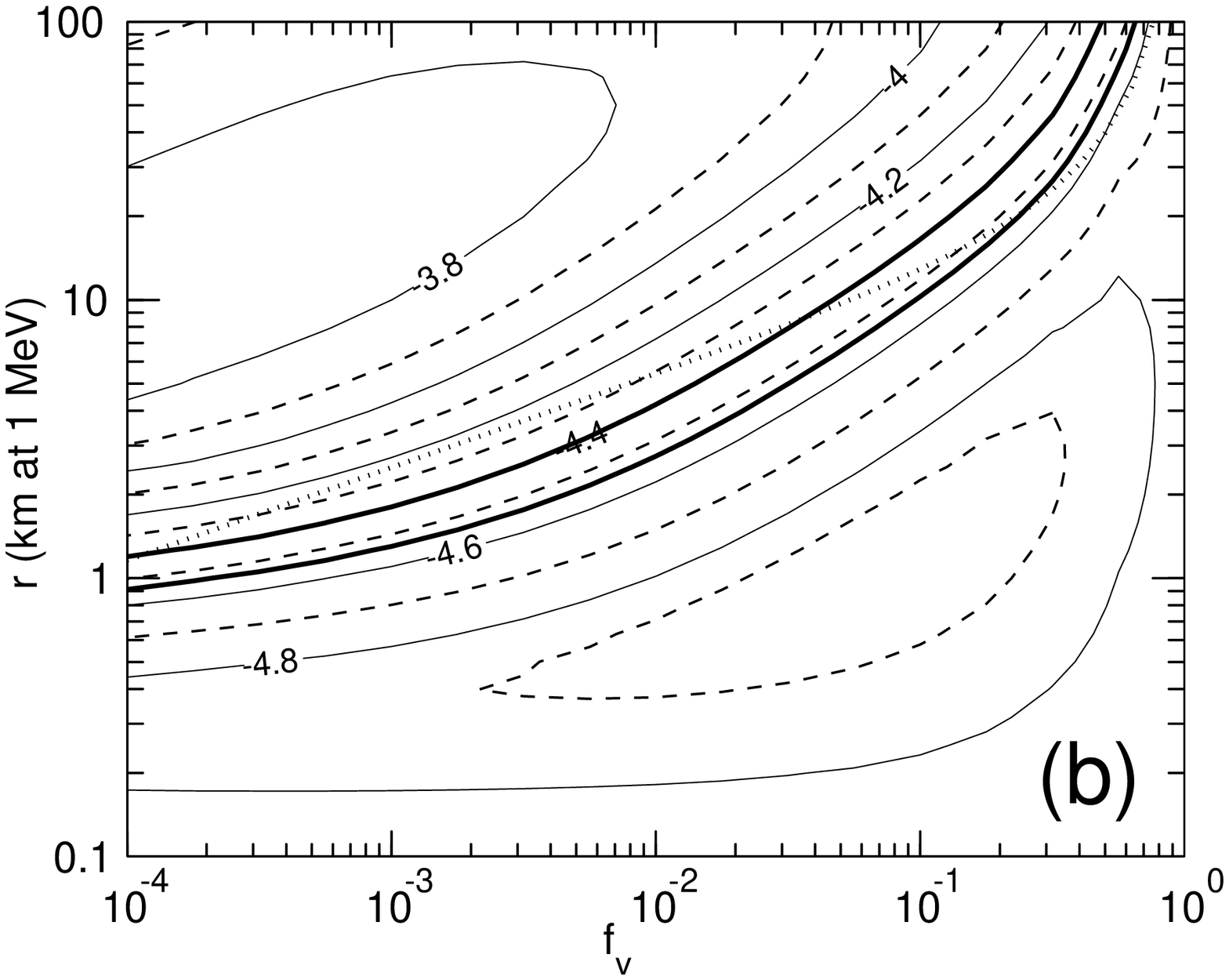}
\epsfysize=6.5cm
\epsffile{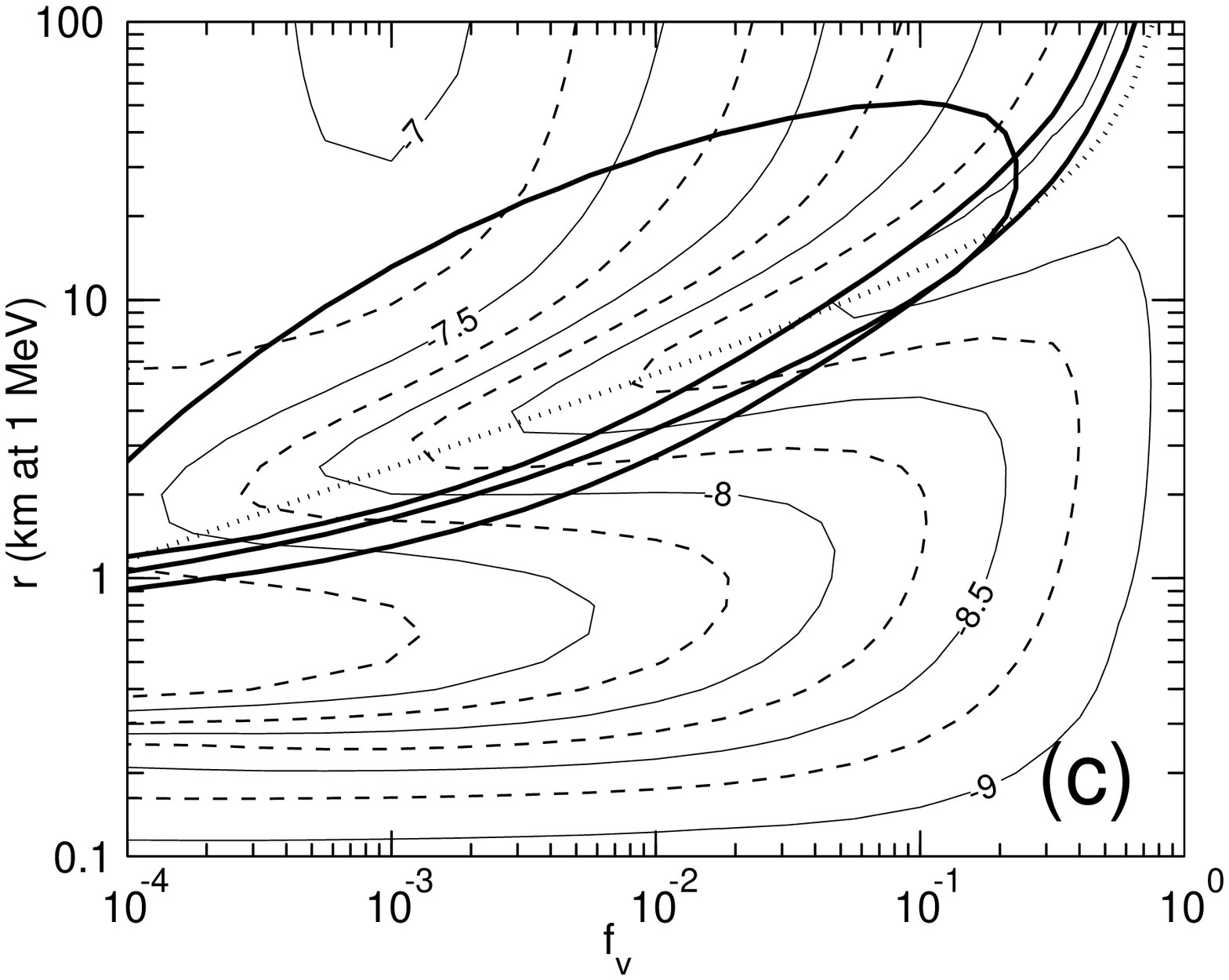}
\vspace{0.2cm}
\caption[a]{\protect The light element yields
(a) $Y_p$,
 (b) $\log_{10}(\DH)$, and (c) $\log_{10}(\ZLiH)$ as a function of the
 high-density volume fraction $f_v$ and the distance scale $r$,
 for the CC (centrally condensed) geometry and $\omega_b = 0.030$.
   The density contrast $R$ is fixed by $f_vR =
 100$.  The observational limits $Y_p \leq 0.248$ and $\DH =
 2.9$--$4.0\times10^{-5}$ are indicated by thick lines (both are shown
 superposed on the $\ZLi$ yield in (c)).  The dotted
 line shows the $r(f_v)$ relation, ``optimum valley'',
 chosen for Fig.~\ref{fig:ccR}.
 The limit of small $r$ or large $f_v$ gives the SBBN result.
} \label{fig:ccr}
\end{figure}

\begin{figure}[tbh]
\epsfysize=6.5cm
\smallskip
\epsffile{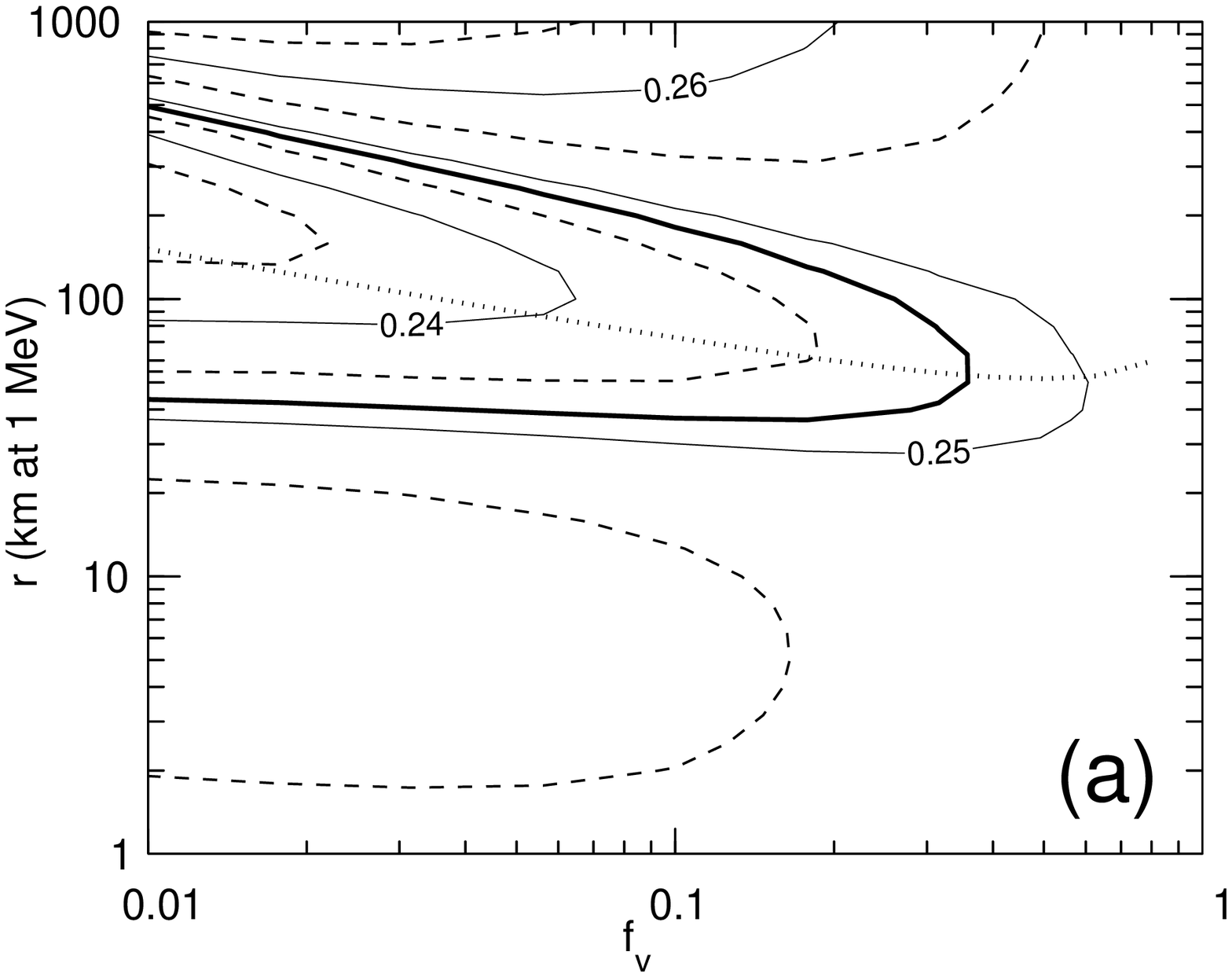}
\epsfysize=6.5cm
\smallskip
\epsffile{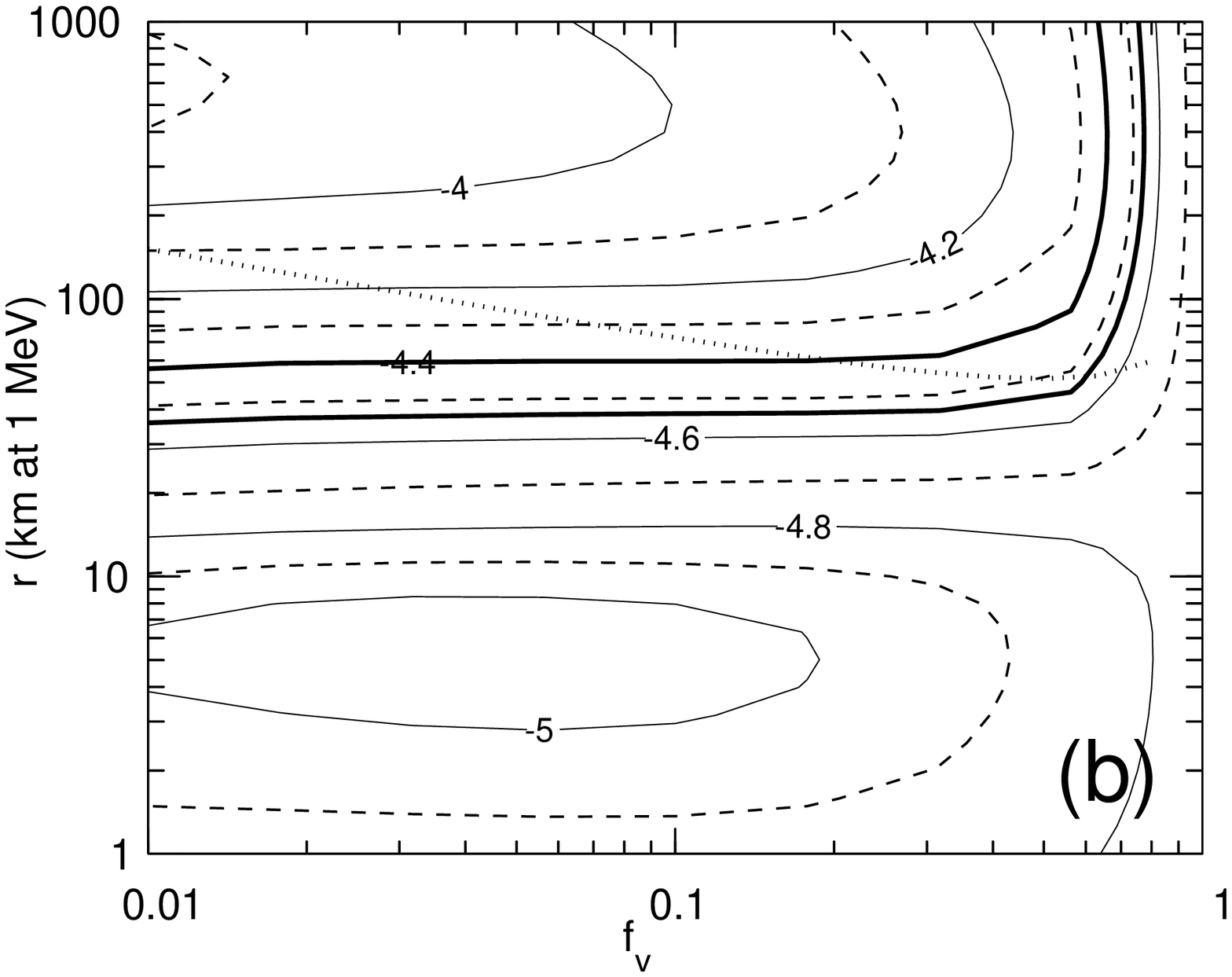}
\epsfysize=6.5cm
\smallskip
\epsffile{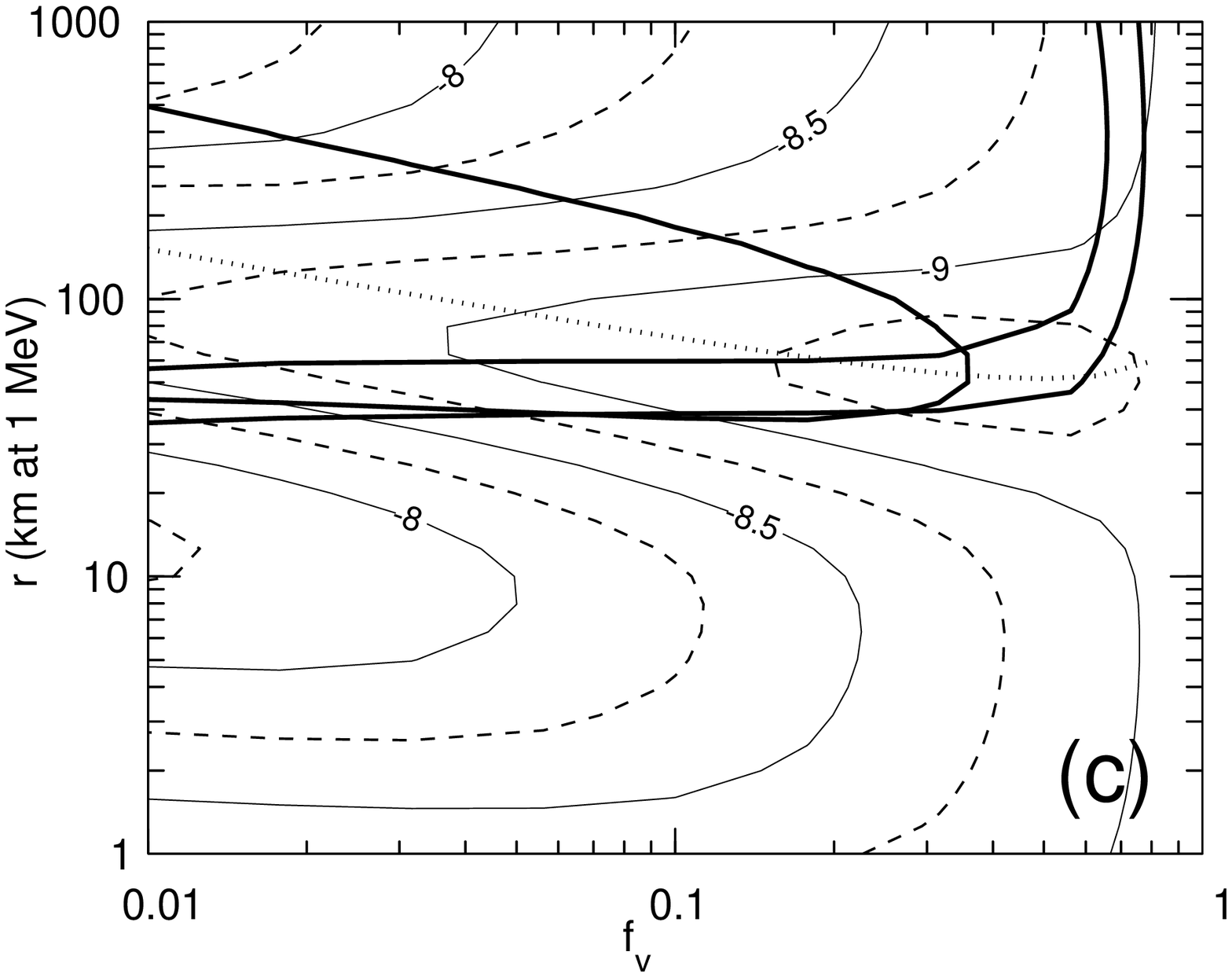}
\vspace*{0.2cm}
\caption[a]{\protect
Same as Fig.~\ref{fig:ccr}, but for the SS (spherical shell)
geometry. The dotted
 line shows the $r(f_v)$ relation, ``optimum valley'',
 chosen for Fig.~\ref{fig:ssR}.
}\label{fig:ssr}
\end{figure}

It has been customary in IBBN work to choose $R$ sufficiently
large to get rid of the $R$-dependence.  This reduces the
dimension of the parameter space by one, and maximizes the effect
of the inhomogeneity.  We first follow this practice by fixing
$f_vR = 100$, and show the IBBN yields of $\UHe$, $\D$, and $\ZLi$
in the remaining two dimensional parameter space $(f_v,r)$ in
Fig.~\ref{fig:ccr} for the CC geometry, and in
Fig.~\ref{fig:ssr} for the SS geometry.

All our distances given in km refer to comoving distance at $T =$ 1 MeV.
Note that
1 km at 1 MeV corresponds to 8.5 m at 100 MeV and
to $1.9\times10^{-4}$ pc today.

We see that IBBN leads to acceptable $\UHe$ and $\D$ yields for
$\omega_b = 0.030$.  This occurs for distance scales which are
near the ``optimum distance scale'' for IBBN.  Moreover, choosing
the ``high'' observed value for $Y_p$, and the ``low'' observed
value for $\DH$ gives us agreement for moderate (not very small)
values of $f_v$, where the $\ZLi$ yield is the lowest.
Choosing a lower $Y_p$ and a higher $\DH$ would move the
agreement towards smaller $f_v$, where the $\ZLi$ yield would be
significantly larger than in SBBN.  However, for the SS geometry,
somewhat higher $\DH$ values would be fine.

For SBBN with $\omega_b = 0.030$,
all three observables, $Y_p$, $\DH$, and $\ZLiH$ are
outside their observed ranges.
While the inhomogeneity at the right distance scale
had the effect of moving $Y_p$ and $\DH$
into the observed range, this is not true for
$\ZLiH$.  For the CC geometry IBBN does not reduce the $\ZLi$
yield appreciably.  For the SS geometry there is a reduction in $\ZLiH$ for
not too small $f_v$, which could arguably be sufficient.

\begin{figure}[tbh]
\epsfysize=7.0cm
\smallskip
\epsffile{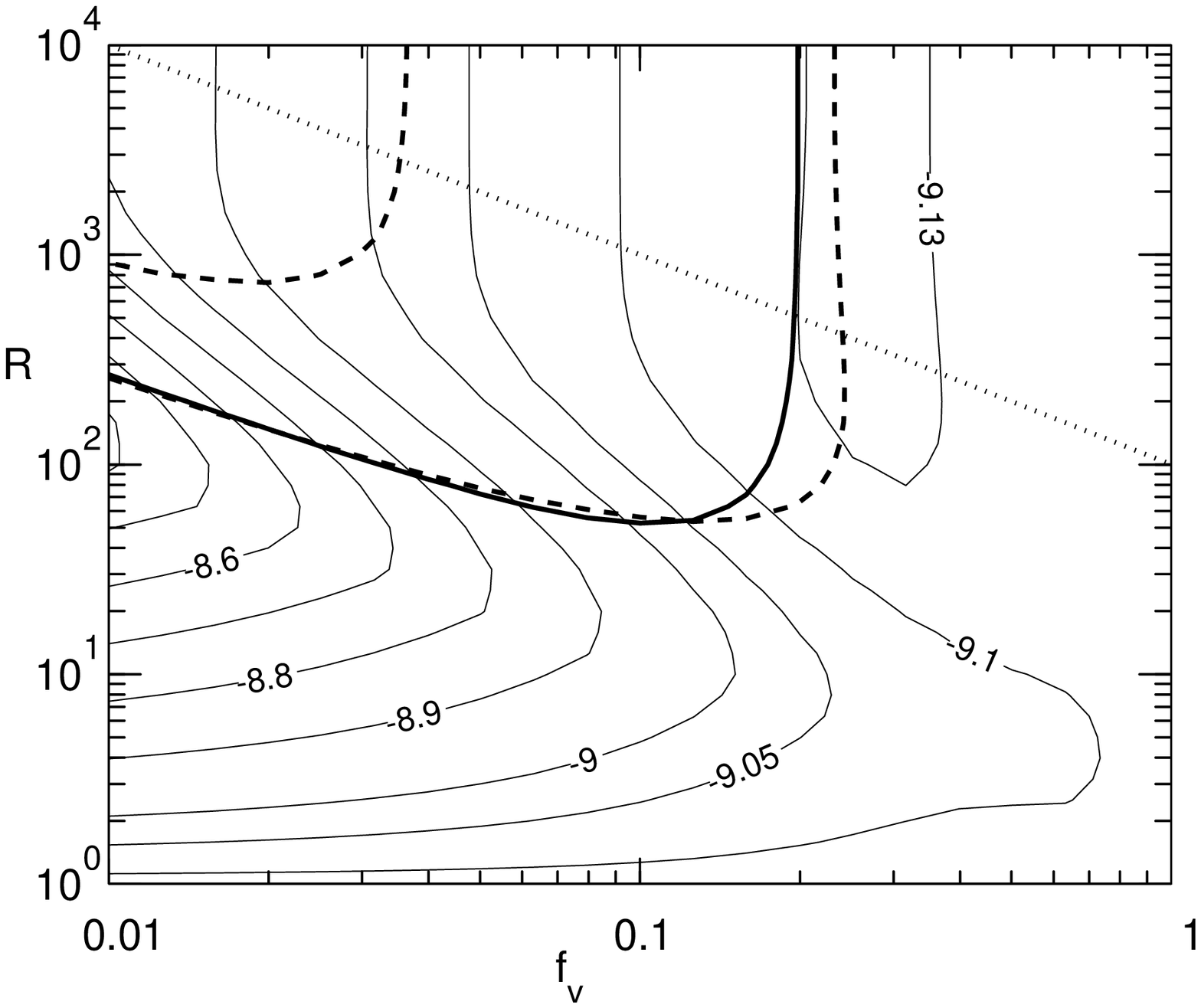}
\vspace*{0.2cm}
\caption[a]{\protect The lithium yield, $\log_{10}(\ZLiH)$, as a function of the
 high-density volume fraction $f_v$ and the density contrast $R$
 for the CC geometry.  The distance scale $r$ corresponds to the
 dotted line in Fig.~\ref{fig:ccr}.
   The observational limits $Y_p \leq 0.248$ (above the solid line) and $\DH =
 2.9$--$4.0\times10^{-5}$ (between the dashed lines) are indicated by thick lines.
 The dotted line corresponds to the relation $f_vR = 100$ used for
 Fig.~\ref{fig:ccr}. Taking $R = 1$ or $f_v = 1$
 gives the SBBN result.  To bring out
 the slight decrease in the$\ZLiH$ yield, we have included the
 $\lgZLiH = -9.13$ contour in this figure.
} \label{fig:ccR}
\end{figure}

\begin{figure}[tbh]
\epsfysize=7.0cm
\smallskip
\epsffile{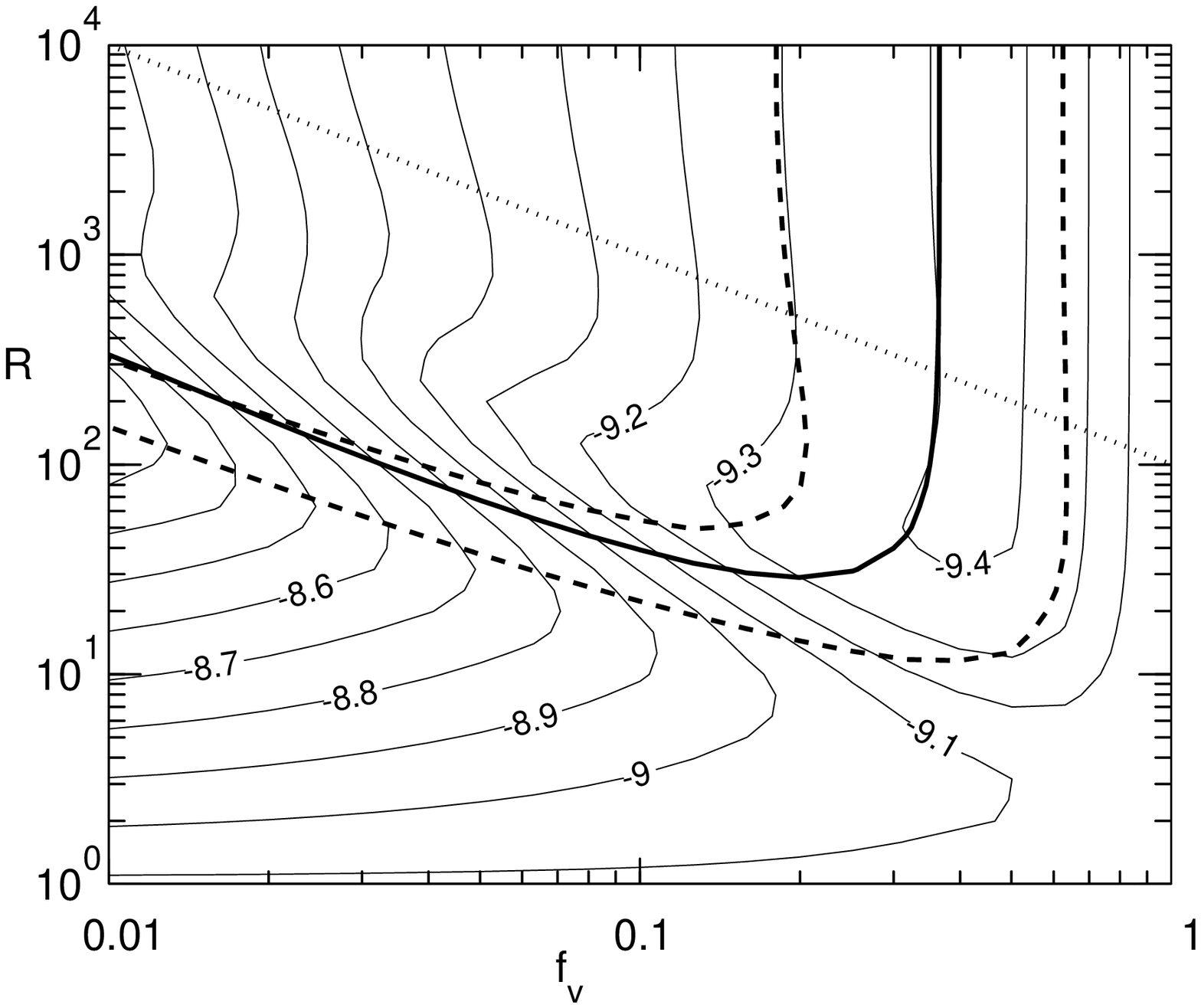}
\vspace{0.2cm}
\caption[a]{\protect Same as
Fig.~\ref{fig:ccR}, but for the SS geometry.
} \label{fig:ssR}
\end{figure}

To see what the effect of the remaining parameter, $R$ has on this
lithium problem, we choose to track the ``lithium valley'' of
Figures \ref{fig:ccr} and \ref{fig:ssr}.  In\cite{KKS99} we
derived the dependence of the optimum scale $\ropt$ on $f_v$ and
$\eta$,
\begin{eqnarray}
    \ropt
        & \propto & {f_v^{1/3}\over (1-f_v)} \eta^{-2/3}
        \qquad \hbox{(CC)}
         \\
    \ropt
        & \propto & {\eta^{-2/3} \over f_v^{1/3}(1-f_v)^{1/3}}
 \qquad \hbox{(SS)}
        . \nonumber
\end{eqnarray}
From
Figures \ref{fig:ccr} and \ref{fig:ssr}, we see that the minimum
$\ZLi$ yield occurs for slightly smaller distance scales than the
minimum $\UHe$ yield.  We choose the proportionality constant so
that $r(f_v)$ tracks the low $\ZLi$ yields.  Thus we have taken
\begin{eqnarray}
    r
        & = & 25 {\rm km} {f_v^{1/3}\over (1-f_v)}
         \qquad \hbox{(CC)}
        \\
    r
        & = & 32.5 {\rm km} {1\over f_v^{1/3}(1-f_v)^{1/3}}
        \qquad \hbox{(SS)}, \nonumber
\end{eqnarray}
(the dotted lines in Figures 1 and 2).
We now allow $R$ to vary independently of $f_v$ and show the
results in Figures \ref{fig:ccR} and \ref{fig:ssR}.

To check that the lithium valley does not shift in $r$ as the
density contrast $R$ is reduced, we repeated this for nearby distance
scales.

These figures quantify the effect of the density contrast $R$ in
IBBN.  We see how the effect indeed saturates for $f_vR \gtrsim 100$.
For IBBN to have the desired effect of bringing $Y_p$ and $\DH$
into agreement with observations for a high baryon density (here
$\omega_b = 0.03$), we need a density contrast of $R \gtrsim 4/f_v$.
Varying $R$ does not have much effect on the lithium problem.
The smallest $\ZLi$ yield we get with acceptable $Y_p$ and $\DH$
is $\ZLiH = 7.4\times10^{-10}$ for the CC geometry and
$\ZLiH = 4.0\times10^{-10}$ for the SS geometry.

We found that IBBN with the CC geometry is not able to reduce the
$\ZLi$ yield significantly, but the SS geometry can reduce it by
about a factor of two.
Thus one could argue for a marginal IBBN case, where we have an
initial inhomogeneity approximating the SS geometry, i.e., the
surface-to-volume ratio of the high-density regions is relatively
large.  With a high-density volume fraction $f_v = 0.2$--$0.5$,
density contrast $R > 30$, and a distance scale $r = 35$--$70$ km
at 1 MeV, we get $Y_p$ near the upper limit 0.248, $\DH$
comfortably in the observed primordial range, and $\ZLiH \sim
4\times10^{-10}$.  With the uncertainties in the $\ZLiH$
observations and BBN reaction rates leading to $\ZLi$ this is in agreement with
observations if one is willing to accept a depletion factor of
about two.

We are, however, not aware of a well-motivated scenario for producing a baryon
inhomogeneity both with this geometry and this distance scale
in the early universe.

A baryon inhomogeneity could be generated during inflation as an
isocurvature perturbation, or after inflation in some phase
transition.  In the latter case, the distance scale must be less
than the horizon size during the phase transition.  At the
electroweak phase transition, the horizon size is about 3 km,
and the expected distance scale is significantly smaller than this,
too small to have a large
effect on BBN\cite{KKS99}.

The QCD transition is the most studied candidate for generating
the inhomogeneity for IBBN\cite{earlyIBBN}.  The horizon size at the QCD transition
is about 2000 km.  In a first-order transition, the
resulting baryon density contrast could well be large enough, and
the distribution of distance scales\cite{Meyer} would be fairly narrow.
In homogeneous thermal nucleation, the distance scale depends
on the latent heat $L$ and the surface tension $\sigma$ of the phase transition.
Typical estimates from lattice QCD calculations\cite{lattice} give
distance scales which are much too small for our scenario, although
combinations of $L$ and $\sigma$ leading to sufficiently large distance scales
may not be completely ruled out\cite{IKKL}.  Heterogeneous
nucleation due to impurities could also lead to larger distance scales, of
about 1 km\cite{Christiansen}.  (Note that in the literature these
distances are often quoted as comoving at 100 MeV, and
thus smaller by a factor of 1/117).

Because of primordial perturbations the nucleation of the phase transition
does not
take place in homogeneous conditions. The cosmic fluid is undergoing
acoustic oscillations at subhorizon scales.  Since the nucleation rate
is extremely sensitive to the temperature, thermal nucleation would occur
only at cold spots.  This could lead to a distance scale of the order
of 1 km\cite{Ignatius}, which is large enough to affect
nucleosynthesis.
However, the inhomogeneities produced by the QCD transition appear
likely
to resemble the CC geometry more than the SS geometry.

\section{Conclusions}

It is intriguing, that the ``high'' baryon density $\omega_b \sim 0.03$,
suggested by
recent CMB results fits so well with the ``low'' $\DH$ and ``high''
$Y_p$ observations in the IBBN scenario.
However, the $\ZLi$ yield is significantly higher than the standard
estimates for its primordial value.

For the SS geometry,
IBBN can reduce the lithium yield by about a factor of two in the
best case, but for $\omega_b = 0.03$, we still get $\ZLiH =
4\times10^{-10}$, requiring depletion by at least a factor of two
in the Spite plateau.
This may be marginally acceptable, especially if one allows for
the uncertainty in the $\ZLi$ yield due to the reaction rates.

We note for comparison, that the widely accepted ``low-D"
determination $\omega_b = 0.020$ in SBBN yields $\ZLiH =
3.4\times10^{-10}$ and thus also requires significant lithium
depletion\cite{BNT00}.

The better developed scenarios for producing the kind of
inhomogeneity needed for IBBN
tend to lead
to the CC type of geometry.  For this geometry, IBBN is not able
to reduce the lithium abundance significantly below the SBBN
result, so that depletion by a factor of four would be needed.
This seems to be considered unacceptable at present.

\section*{Acknowledgements}

We thank
the Center for Scientific Computing (Finland) for computational resources.

\end{document}